# Impact of IPv4-IPv6 Coexistence in Cloud Virtualization Environment

Mohammad Aazam[1], Eui-Nam Huh[2]
Innovative Cloud and Security Lab, Department of Computer Engineering
Kyung Hee University, Suwon, South Korea.
[1]aazam@ieee.org
[2]johnhuh@khu.ac.kr

***Abstract*— Since January 2011, IPv4 address space has exhausted and IPv6 is taking up the place as successor. Coexistence of IPv4 and IPv6 bears problem of incompatibility, as IPv6 and IPv4 headers are different from each other, thus, cannot interoperate with each other directly. The IPv6 transitioning techniques are still not mature, causing hindrance in the deployment of IPv6 and development of next generation Internet. Until IPv6 completely takes over from IPv4, they will both coexist. For IPv4-IPv6 coexistence, three solutions are possible: a) making every device dual stack, b) translation, c) tunneling. Tunneling stands out as the best possible solution. Among the IPv6 tunneling techniques, this paper evaluates the impact of three recent IPv6 tunneling techniques: 6to4, Teredo, and ISATAP, in cloud virtualization environment. In virtual networks, these protocols were implemented on Microsoft Windows (MS Windows 7 and MS Windows Server 2008) and Linux operating system. Each protocol was implemented on the virtual network. UDP audio streaming, video streaming and ICMP-ping traffic was run. Multiple runs of traffic were routed over the setup for each protocol. The average of the data was taken to generate graphs and final results. The performance of these tunneling techniques has been evaluated on eight parameters, namely: throughput, end to end delay (E2ED), jitter, round trip time (RTT), tunneling overhead, tunnel setup delay, query delay, and auxiliary devices required. This evaluation shows the impact of IPv4-IPv6 coexistence in virtualization environment for cloud computing.**

***Keywords—virtualization; IPv4-IPv6 coexistence; IPv6 tunneling; 6to4; Teredo; ISATAP***

## I. INTRODUCTION

With today's ubiquitous way of usage of IP addresses and IP based devices, IPv4 is facing many problems, among them, address shortage is the most important and urgent one. After the exhaustion of IPv4 address space, IPv6 is now formally taking over. But IP nodes cannot be upgraded to IPv6 at once. It is a time taking process and IPv4 and IPv6 coexistence has been under discussion from the past decade [1]. Keeping in view the importance of IPv6, the transition from IPv4 to IPv6 has started [1] [10]. To make this transition manageable and easy, a lot of research has been done. It is not practical to change the Internet or any large network altogether into an IPv6 network at once [1] [6]. All the routers cannot be upgraded simultaneously. Hence transition is only feasible in a gradual manner due to complexity of the Internet and huge number of devices accessing it [19]. Also, there will be islands of both IPv4 and IPv6 networks [7]. For example, two IPv6 devices want to communicate with each other but there is an IPv4 network between them, then IPv6 packets have to be routed from that IPv4 network. Network layer virtualization provides a very useful mean to achieve the coexistence of both these versions of IP, described more in section II. As IPv4 and IPv6 headers are different from each other in terms of the header format (fields, the address format and address size are different), so some mechanism is always required that makes it possible to route the packet between different islands, i.e. the IPv4 island and the IPv6 island. For this purpose, three mechanisms: dual stack, translation, and tunneling, are available in the literature [11].

Dual stack means nodes (routers and other IP devices) can understand both IPv4 and IPv6 packets and translate between each other and make communication seamless. To achieve this, the router or node should be enhanced to a dual stack node, which may not be easily possible all the time because of upgrading cost and the cost of time.

Translation means direct conversion of the header fields of protocols (e.g., IPv4 into IPv6 and vice versa). It may include transformation of the protocol headers and payload. Translation bears the problem of feature loss, when there is no clear mapping available for the protocol being translated into another one. In case of IPv6, when translated into IPv4 header, it will lead to the loss of IPv6 flow label field, as there is no mapping available for the 20-bit flow label field in IPv4 header.

Tunneling is a technique in which one protocol is encapsulated in another protocol, according to the network where the packet is to be routed [5], [13]. In case of IPv6 tunneling, if an IPv6 source communicating with an IPv6 destination and an IPv4 network is between them, then IPv6 packets must be tunneled into the IPv4 header so that the IPv6 packet gets routed through the IPv4 network and reaches its intended IPv6 destination. Figure 1 depicts IPv6 tunneling.

As translation bears the problem of feature loss, so it is not always a feasible solution, specially on a large scale. It is used only when two incompatible nodes have to communicate directly. Similarly, it is also not feasible to make every device dual stack. Specially on a large scale, it will not only cost a lot, but also, it is very time consuming and effort taking. The only feasible way to let IPv4 and IPv6 work in coexistence is to do tunneling, as it avoids complexities [1]. It requires a bit of configuration on the end nodes, border routers, and DNS server. Intermediate network or the whole IPv4 Internet remains absolutely unchanged. With the passage of time, IPv4-only nodes would be upgraded or replaced by IPv6 enabled nodes and by this; migration towards IPv6 would be completed.

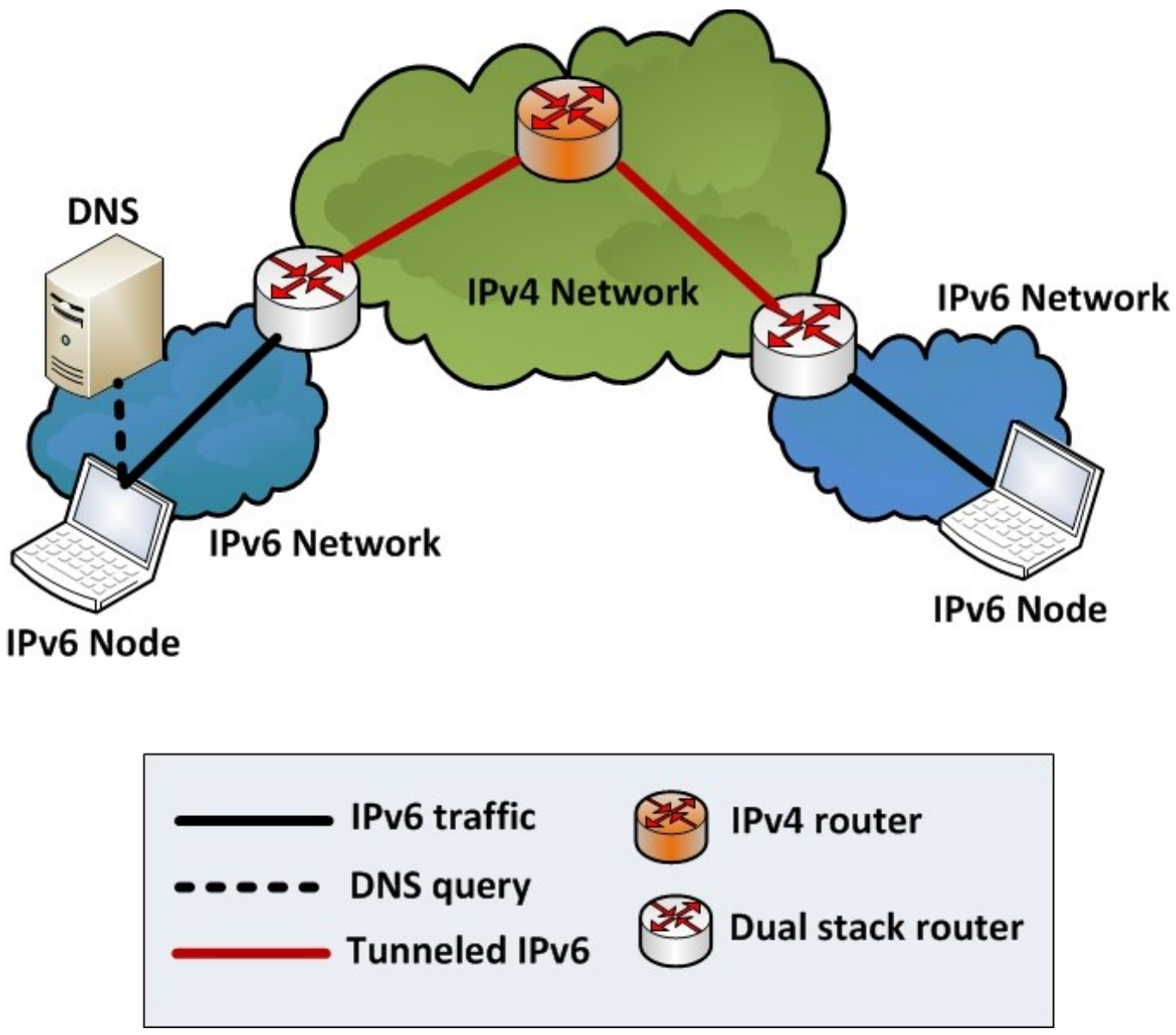

Figure 1 IPv6 tunneling

## II. IPv6 Tunneling Techniques

This research activity covers recent IPv6 tunneling techniques, named: 6to4, Teredo, and ISATAP. These techniques are not only recent, but also they have been standardized by the IETF with their individual RFCs. They have been implemented in all the most widely used operating systems.
These techniques have been described below:

*A. 6to4:* 6to4 is an automatic way to connect isolated IPv6 sites (domains/hosts), attached to an IPv4 network that has no native IPv6 support. 6to4 Relay is provided for such IPv6 sites to visit IPv6 native network before they can obtain native IPv6 connectivity [8]. With 6to4, the current IPv4 network is treated as the link layer and the existing IPv4 routing infrastructure is utilized to forward IPv6-in-IPv4 encapsulated packet. The host in 6to4 site uses a 6to4 IPv6 address (2002:IPv4Address::/16) as the communication identifier [17]. When the IPv6 packet goes through the 6to4 router, the IPv4 address of tunnel end point can be found within the 6to4 address, then a tunnel is formed without explicit configuration. 6to4 is designed only as a temporary mechanism and it will be replaced in future by other mechanisms using permanent IPv6 addresses [10].

*B. Teredo:* Most of the NAT devices does not allow direct encapsulation of IPv6 in IPv4 header. Teredo provides a service that enables nodes located behind one or more IPv4 NATs to obtain IPv6 connectivity by tunneling packets over UDP. With Teredo, the current IPv4 network is treated as the link layer and the existing IPv4 routing mechanism is utilized to forward IPv6-in-UDP-in-IPv4 encapsulated packets, shown in figure 2. Teredo host first gets an IPv6 prefix from the Teredo server, then an IPv6 address is formed with special format (Prefix : Server IPv4 : Flags : Port : Client IPv4) [21]. The communication between Teredo hosts can be made directly with an IPv6-in-UDP-in-IPv4 tunnel. The connectivity to IPv6 native network will be achieved with the Teredo relay gateway. The automatic tunnels between Teredo hosts distribute the traffic between them and share the burden of Teredo relay gateway.

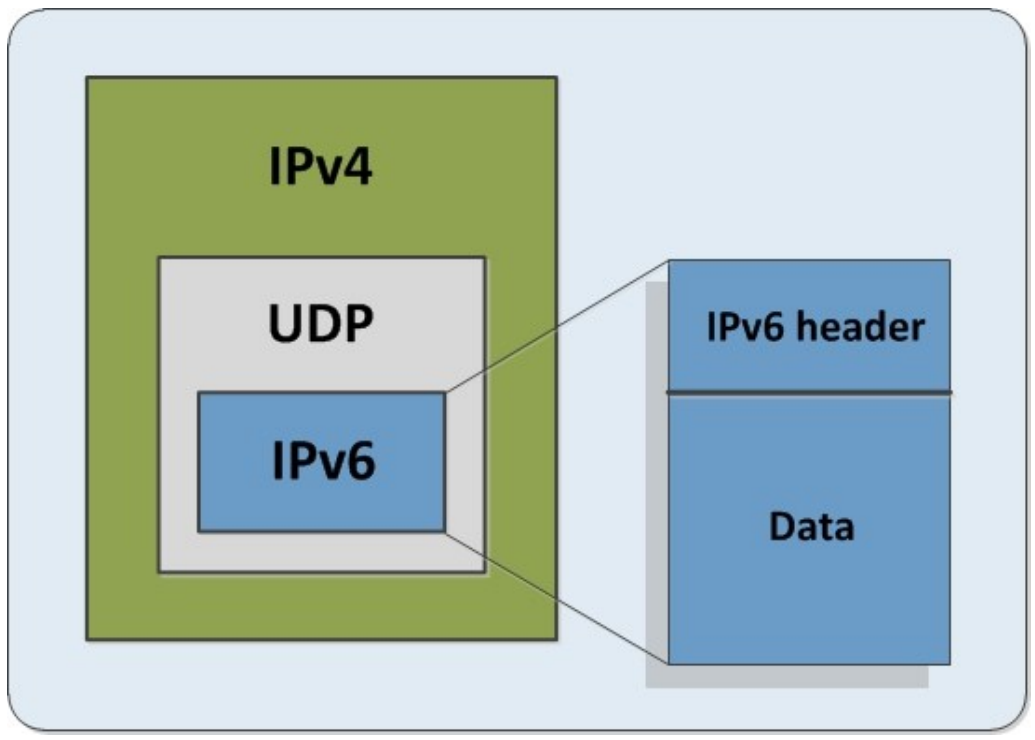

Figure 2 Teredo tunneled IPv6

*C. Intra-Site Automatic Tunnel Addressing Protocol (ISATAP):* ISATAP is designed for the intra-site scope. ISATAP connects isolated IPv6 nodes within IPv4 sites via automatic IPv6-in-IPv4 tunnels. That is why it is called intra-site [9]. With ISATAP, the intra-site IPv4 network is viewed as a link layer for IPv6, and other nodes in the intra-site network are viewed as potential IPv6 hosts/routers. An automatic tunneling abstraction is supported, which is similar to the Non-Broadcast Multiple Access (NBMA) model. An ISATAP host gets a 64-bit prefix from the ISATAP server. Then an ISATAP address is formed with its own interface identifier (::0:5EFE:IPv4Address). After that, the ISATAP hosts can connect with each other via the IPv6-in-IPv4 tunnel with ISATAP addresses [16]. Furthermore, ISATAP can be used to provide connectivity to the outside IPv6 network together with other transition mechanisms. For example, if the site gateway is supported with 6to4 and holds the 6to4 prefix as an ISATAP server and the IPv6 hosts among this site can use ISATAP to get intra/inter-site IPv6 connectivity.

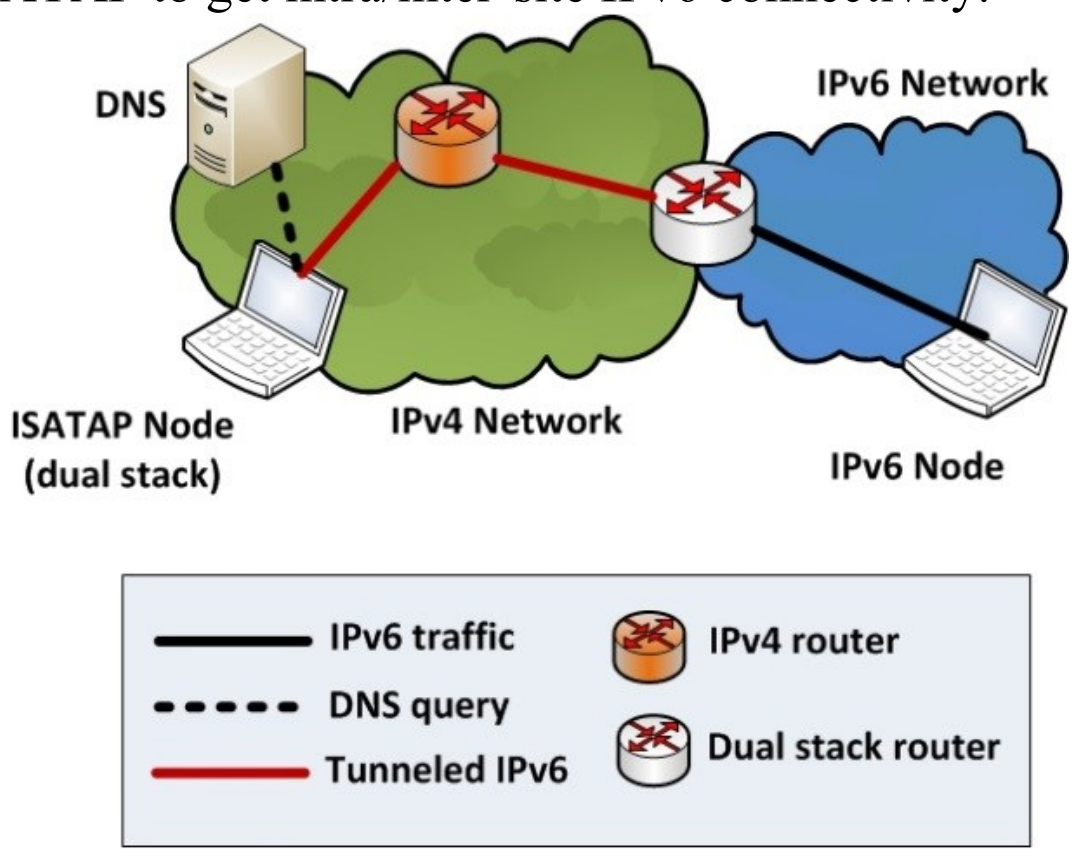

Figure 3 ISATAP tunneled IPv6

## III. Virtualization Environment Setup

Today, the Internet has changed, with the facility of ubiquitous computing. With cloud computing, computing power, storage, applications, and application development is now being outsourced [4]. Virtualization plays a vital role in cloud computing, since this is the key component cloud computing is built upon [2]. For the optimization of resource sharing and utilization on the cloud, virtualization [3] is a feasible way to achieve IPv4-IPv6 coexistence [1]. IPv4 and IPv6 must be treated independently for addressing, routing, as well as for operations, administration, and maintenance (OAM). To achieve this, the required level of isolation is provided by network layer virtualization [1]. IEEE 802.1Q is a mean to have separate LANs over a single bridged LAN, through virtualization. Since virtualization is among the key components of cloud computing, so it will be effective to create IPv4-IPv6 coexisting environment through virtualization. But, on the other hand, it is also important to know that when virtual machines and virtual LANs are addressing users' requests, how much this IPv4-IPv6 coexistence is going to impact. The biggest challenge that has been faced is that which tunneling technique is to be deployed in what circumstances and in which scenario [20]. This study shows the tunneling overhead and traffic comparison of the most widely deployed IPv6 tunneling techniques in cloud virtualization environment, extending [24][25]. This study shows that to what extent IPv4-IPv6 coexistence impact on virtualization in cloud computing.

### A. *Virtual Network Setup for IPv4-IPv6 Coexistence*

Since this study is based on IPv4-IPv6 coexistence, so hybrid IPv4-IPv6 virtual network was needed to create the need for tunneling and then evaluate the performance of these tunneling protocols. IPv6 virtual networks and IPv4 virtual networks were created in such a way that they make a hybrid network. The overall topology was divided into three parts: the IPv4 Internet, IPv6 network, and IPv4 network. The IPv4 Internet provides transit service to IPv4 traffic and IPv6 traffic. The IPv6 traffic is routed by encapsulating IPv6 packets inside IPv4. It uses dual stack nodes at the edges of the infrastructure, overlapping both IPv4 network and IPv6 network. The IPv4 network carries native IPv4 traffic, while the IPv6 network carries native IPv6 traffic, generated by the IPv6 end users. The nodes involved in the topology include hosts, DNS server, and routers. For Teredo, one NAT box was also configured to fulfill the recommendation for Teredo setup. Teredo can work without the involvement of NAT device, but then in that case, it would be useless, because the actual reason to use Teredo is to allow IPv6 encapsulated inside IPv4 to pass through NAT. As, Teredo is being designed to provide end-to-end tunneling to the nodes residing in private IPv4 network, so in this research activity, this thing was considered. Figure 4 depicts the general setup more comprehensively. Teredo's setup is shown in figure 5.

ISATAP works where one node (dual stack ISATAP node) is residing inside an IPv4 network and it has to communicate to an IPv6-only node, sitting inside an IPv6 network [9]. ISATAP setup was in such a way that communicating node, the ISATAP node, as shown in the figure 3, is a dual stack node, residing in an IPv4 network. This node needs to communicate to an IPv6 node, residing in an IPv6 network. Now, the traffic of and from both ISATAP node and IPv6 node has to go through IPv4 network. Those IPv6 packets have to be routed through IPv4 network, which is possible with ISATAP tunnel. By default, the IPv6 protocol for Windows XP Professional with SP2, Windows 7, Windows Server 2003 with SP1, Standard Edition, and Windows Server 2008 configures a link-local ISATAP address on the Automatic Tunneling Pseudo-Interface for each IPv4 address assigned to a computer. To configure global ISATAP addresses, or to communicate beyond the logical subnet defined by the IPv4 intranet, an ISATAP router is needed [12].

ISATAP router performs two functions: a). advertises its presence and address prefixes, enabling global ISATAP addresses to be configured. b). optionally forwards IPv6 packets between ISATAP hosts on the IPv4 intranet and IPv6 hosts beyond it [14]. An ISATAP router is typically configured to perform both functions, but can perform either individually. Most often, an ISATAP router acts as the forwarder between ISATAP hosts on an IPv4 intranet and IPv6 hosts on an IPv6-enabled portion of an intranet [18]. To demonstrate the use of an ISATAP router between IPv6 and IPv4 intranets, first the setup was separated into a portion that has IPv4 and IPv6 connectivity and another that has IPv4 connectivity only [15]. For ISATAP setup, the ISATAP dual stack node is residing inside VN2 (IPv4), while the receiver node is IPv6-only, residing in VN3 (IPv6). The router which has an incoming interface to VN2 and outgoing interface to VN3 was configured as an ISATAP router, to perform ISATAP tunneling for the ISATAP node and IPv6-only node. Shown in figure 4, this emulates an intranet in which a portion is IPv6 enabled (VN 1) and a portion is not (VN 2 and VN 3).

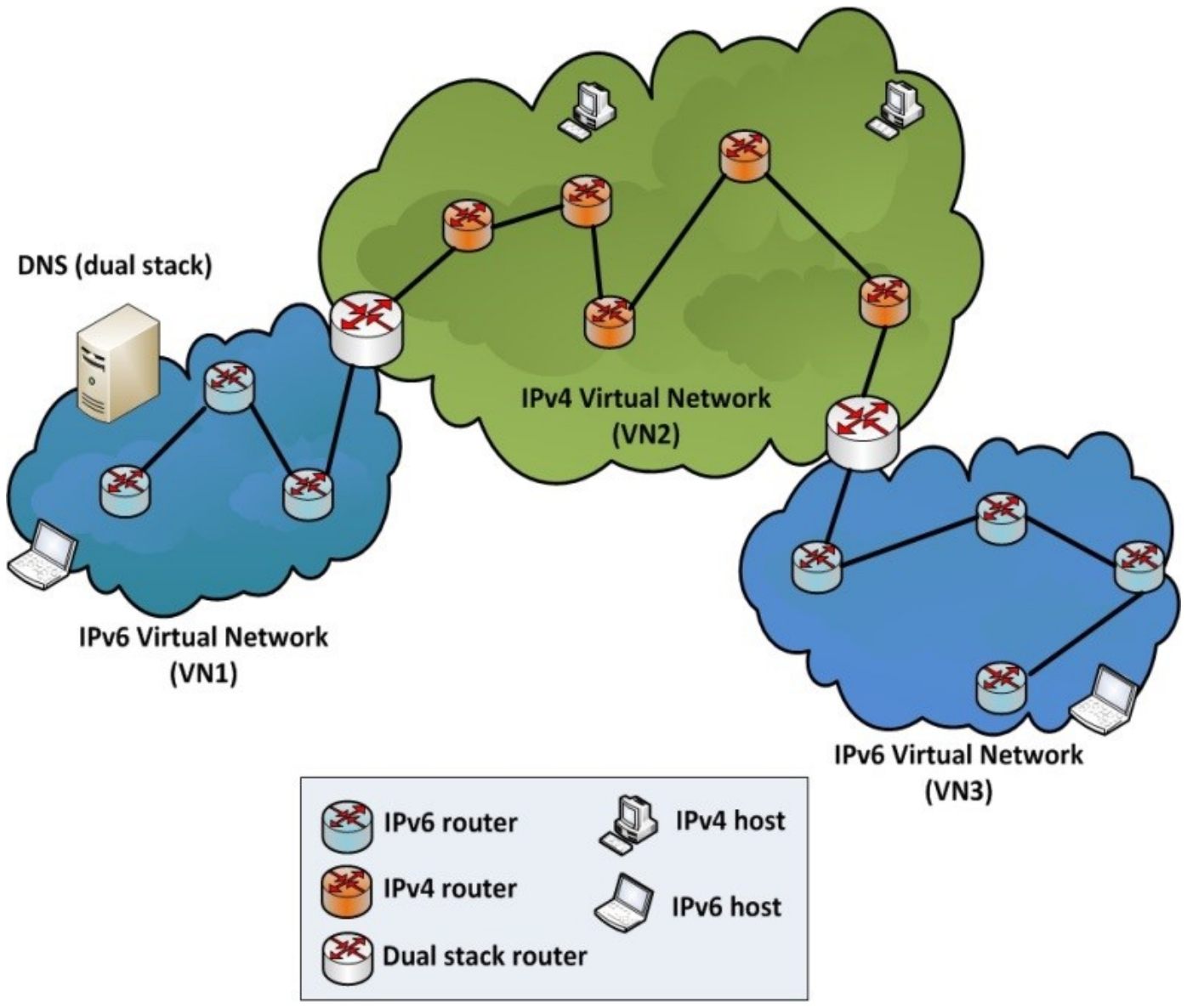

Figure 4 Virtual network setup

In case of Teredo, the host, known as Teredo host was required to be behind NAT device. Teredo tunneling is workable even without putting Teredo client behind NAT. But, Teredo was introduced to serve as a mean for IPv4-IPv6 coexistence for the nodes residing behind NAT [22]. That is why; the recommended scenario was configured in this activity. So, one NAT box was also configured to setup the required scenario for Teredo [23]. Rest of the setup remained physically unchanged, shown in figure 5. Sending node was configured as Teredo client, while receiving node was configured as an IPv6 node. Border router, interfacing IPv6 network from the IPv4 network was configured as Teredo relay router, so that it is able to understand Teredo tunneled packtets and perform dual encapsulation and decapsulation for the clients. The entire configuration was done on Linux (Fedora), through set of commands available in miredo package [23].

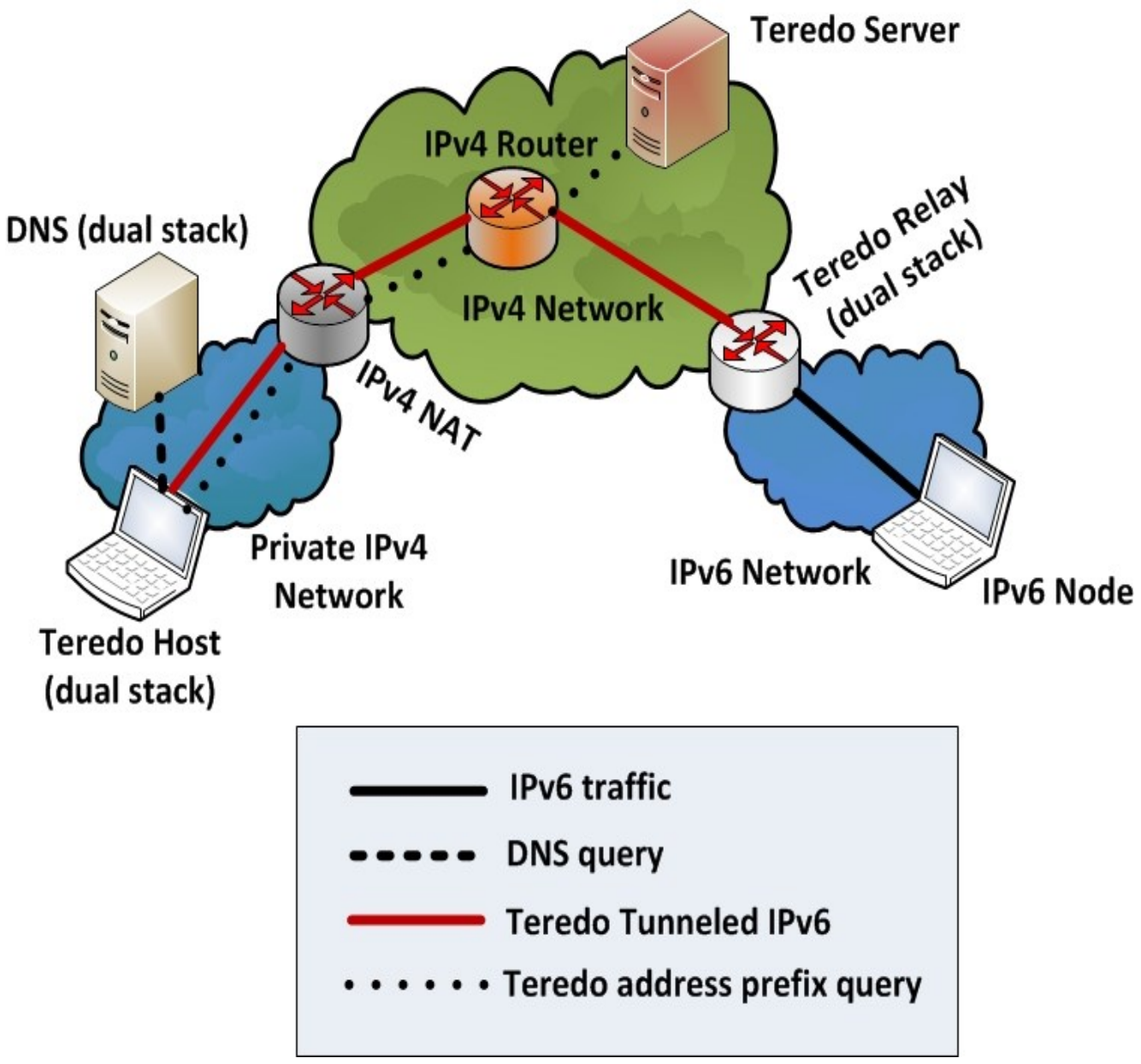

Figure 5. Teredo virtual network setup

6to4, like ISATAP, also requires two routers, but configured as 6to4 routers and capable of providing 6to4 tunnel address prefix to the client asking for that. The dual host was configured as 6to4 host, so that it could ask for 6to4 prefix from the nearest 6to4 router. 6to4 requires standalone 6to4 server when deployed over the Internet or on a large network [17].

## IV. Performance Evaluation and Impact

The performance was evaluated based on certain parameters. The parameters were selected keeping in view the attributes related to tunneling and network layer or routing. Throughput, round trip time, jitter, end to end delay, tunneling overhead, tunnel setup delay, query delay, and auxiliary device required were the parameters used to evaluate the performance.

### A. *Throughput:*

It describes the overall throughput of the protocols. i.e. number of packets received per second.
Formula to calculate throughput is:

$$Throughput_{pps} = No_of_pkts_rcvd / timestamp_{sec}\ pps \qquad (1)$$

Where pps is packets per second.
UDP audio streaming traffic was run end to end to calculate the throughput. There were multiple runs of traffic, whose average was calculated and the final result, as shown in figure 6, was generated.

The packet size remained 1500 bytes. Traffic was run for about five minutes over the network setup. The graph shows that ISATAP performs better in respect of throughput. Its throughput remained up to 45 packets per second. 6to4's throughput remained 35-38 packets per second. Teredo's throughput was around 30-35 packets per second. The reason why ISATAP comes out to be the most efficient one, is that it has to deal with a relatively shorter communication identifier, as compared to 6to4. In case of Teredo, dual encapsulation makes it relatively less efficient.
The average throughput in Kbps comes out as:

$$Throughput = \frac{\left(\frac{P_r}{t_s}\right) * 1500 * 8}{1000}\ Kbps \qquad (2)$$

Where $P_r$ represents number of received packets. $t_s$ represents time in seconds of the received packets. As mentioned before, 1500 was the packet size in bytes, so $\frac{P_r}{t_s}$ is then multiplied with 1500 and then 8 to get the result in bits. Finally, the resultant outcome is divided by 1000 get the final result in Kbps.
As shown in the table 1, ISATAP has the edge of around 11 Kbps on 6to4. ISATAP supersedes Teredo with a difference of more than 40 Kbps.

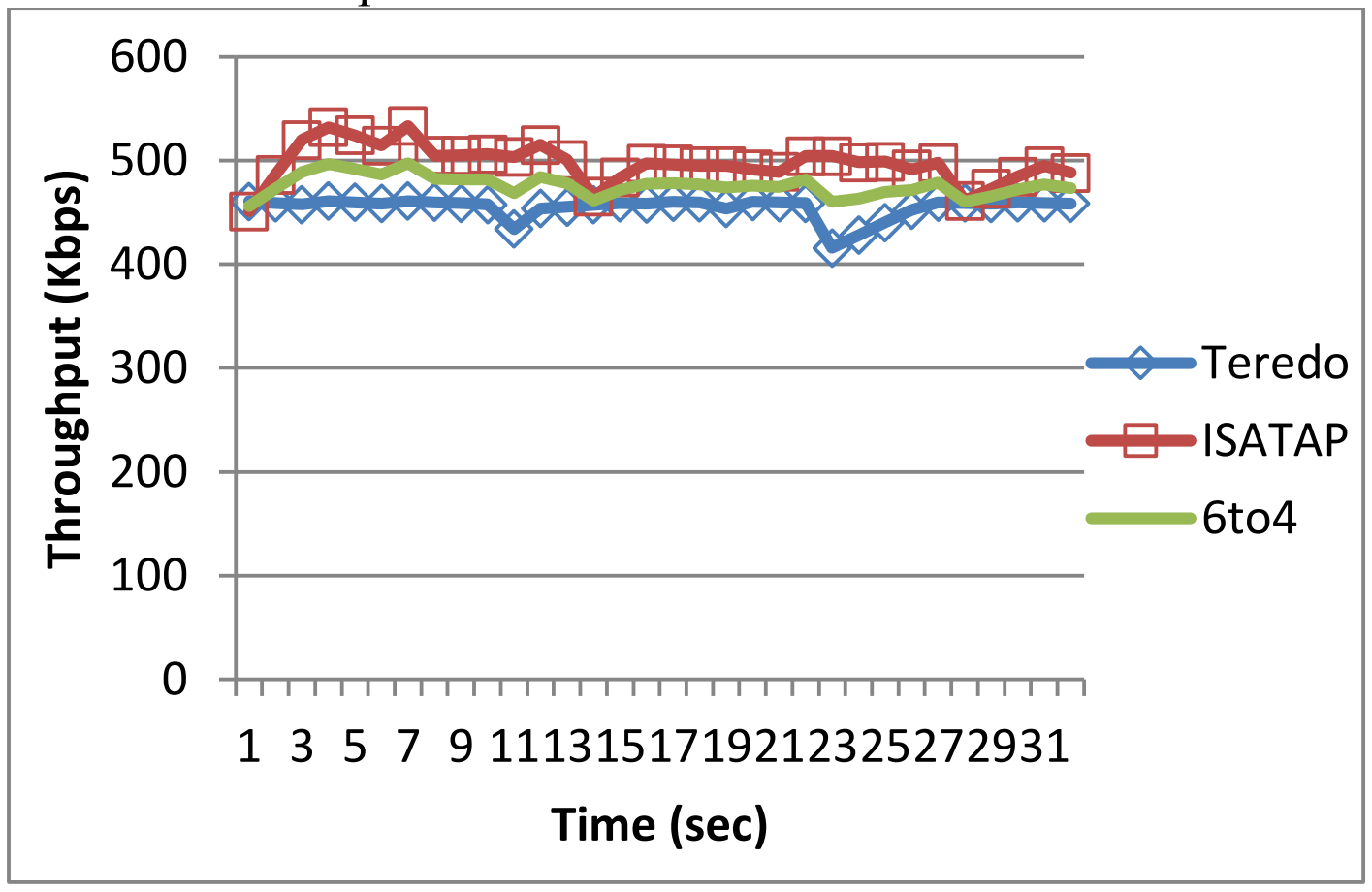

Figure 6 6to4-Teredo-ISATAP throughput (in Kbps)

Average throughput is shown in table 1.

TABLE I. AVERAGE THROUGHPUT

| Protocol | 6to4 | Teredo | ISATAP |
|---|---|---|---|
| Average Throughput (Kbps) | 486.40 | 454.76 | 497.02 |

*B. End to End Delay (E2ED):*It describes delay in milliseconds the traffic incurs, from source node to destinatin node.

The equation for end to end delay is:

$$E2ED = \sum_{i=0}^{n} Tr_i - Ts_i \text{ milliseconds (3)}$$

Where $Tr_i$ represents timestamp of received *i*-th packet and $Ts_i$ represents timestamp of sent *i-th* packet. E2ED is represented in milliseconds here.

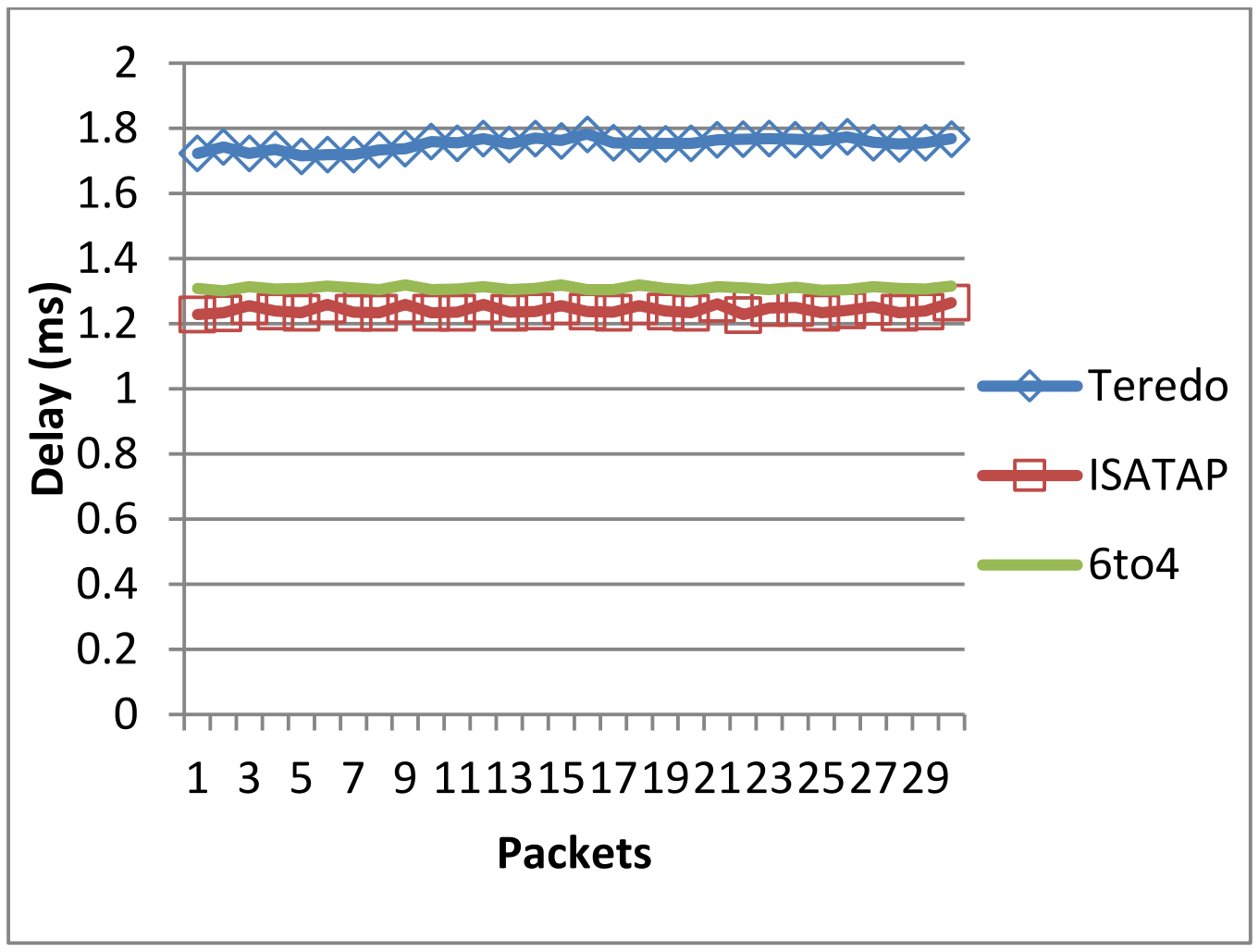

Figure. 7 6to4-Teredo-ISATAP end to end delay

Delay was also calculated on UDP audio streams. It was calculated by subtracting the timestamp of packet received at the receiver end with the timestamp of packet sent by the sender. Again, in terms of delay, ISATAP has got the edge. Its performance is better than 6to4 and Teredo. Teredo traffic incurs more delay, as compared to the other two contenders. Because of the larger communication identifier that 6to4 has to deal with, more delay is incurred, as compared to ISATAP. As discussed before, the host in 6to4 site uses a 6to4 IPv6 address (2002:IPv4Address::/16) as the communication identifier. While for Teredo, the dual encapsulation and decapsulation mechanism consumes more time for processing, thence incurring more delay.

Average delay is shown in the following table:

TABLE II. AVERAGE END TO END DELAY

| Protocol | 6to4 | Teredo | ISATAP |
|---|---|---|---|
| Average Delay (ms) | 1.3103 | 1.7517 | 1.2427 |

Table 2 shows that ISATAP incurs least amount of delay in comparison with 6to4 and Teredo. When these techniques are deployed over the Internet by the service providers, there would be lot more devices each packet has to travel through. So, it is anticipated that this difference would be lot more then. The results in this activity show that in virtualization, what is the effect of tunneling and IPv4-IPv6 coexistence. This difference in performances for each protocol suggests that in wide area network, the performance difference of ISATAP, Teredo, and 6to4 would of this much ratio, most probably.

*C. Jitter:* The variation in delay, jitter, was calculated on the basis of the end to end delay. The formula is given in equation 4.

$$Jitter = \sum_{i=1}^{n} D_i - D_{i-1} \text{ milliseconds} \quad (4)$$

Where *D* represents the delay of packet *i* in milliseconds. The first packet has no previous packet, so jitter for i=0 will be 0.

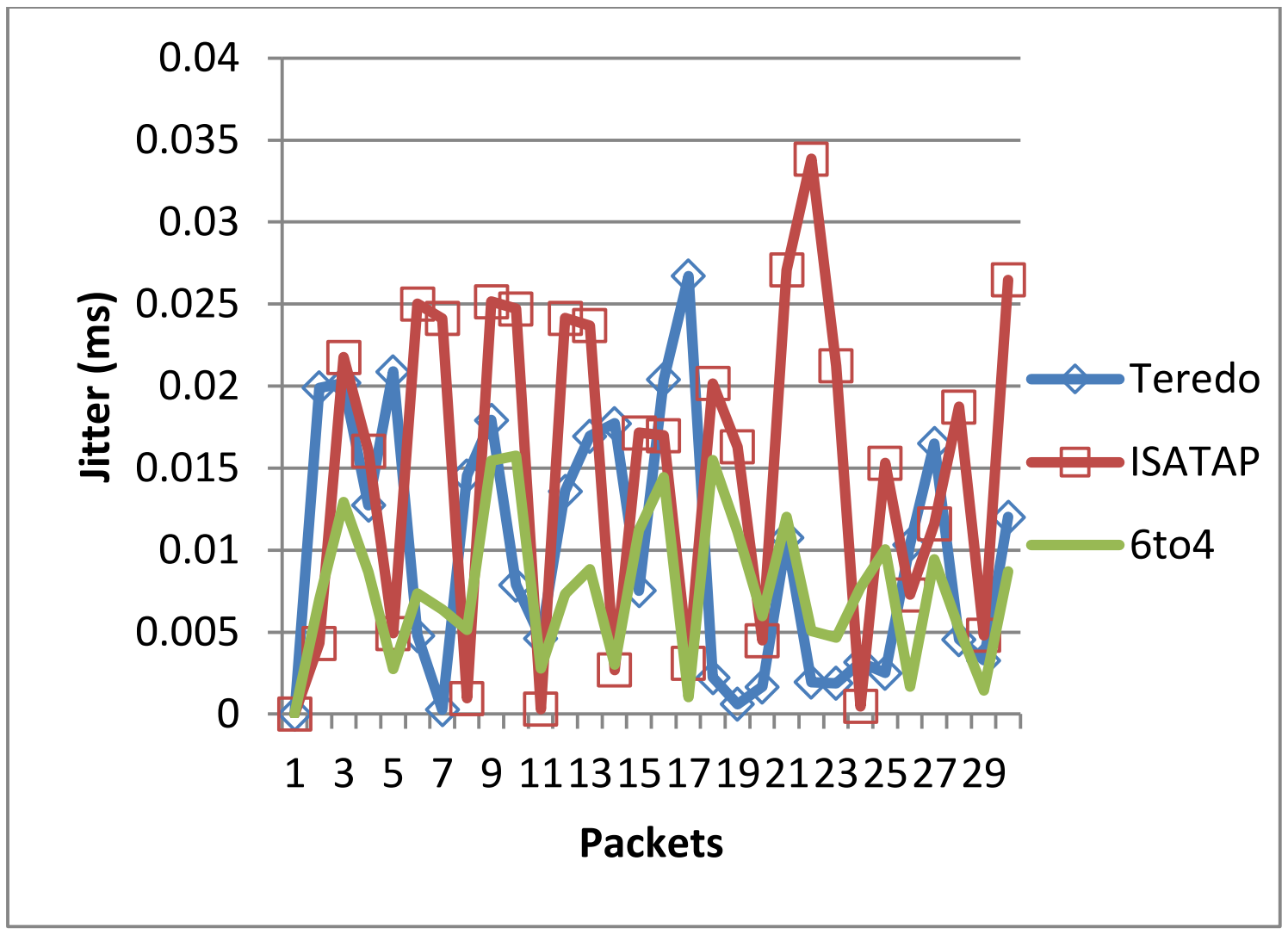

Figure. 8 6to4-Teredo-ISATAP jitter

Jitter is the only parameter in which ISATAP's performance is inferior from 6to4 and Teredo. Teredo being the best in this case. This is because in ISATAP's tunnel refresh packets; which are meant to refresh/maintain the tunnels; are send more frequently. They are sent after every 13 data packets. When the tunnel is being refreshed and maintained, the data traffic is halted for that while. In 6to4, tunnel refresh packets are sent randomly, but less frequently; after around 18-19 data packets. This is the reason for 6to4 being less jittery than ISATAP. In case of Teredo, tunnel refresh packets are sent after every 21 data packets, making it least jittery. This result shows that when real time streaming traffic has to be tunneled through IPv4 network, then Teredo is the best option.

Average jitter is shown in the following table:

TABLE III. AVERAGE JITTER

| Protocol | 6to4 | Teredo | ISATAP |
|---|---|---|---|
| **Average Jitter (ms)** | 0.0225 | 0.0080 | 0.0300 |

The difference revealed from table III shows that when these techniques would be deployed over the Internet, ISATAP tunneled traffic would be containing even more jitter, because when there would be lot more devices involved and the network conditions would also be unpredictable, then those refresh packets would take more time to travel through the network between the client and the ISATAP server. Thence, incurring more jitter and also, affecting the bandwidth as well. Teredo would then be highly appreciated for real-time data traffic.

To elaborate more the performance of these three techniques, based in jitter, variance and standard deviation is also calculated.

- Variance

Variance is the average of squared differences from the Mean. Formula to calculate variance is given in equation 5.

$$S^2 = \sum_{i=1}^{N} \frac{(x_i - M)^2}{N} \qquad (5)$$

Where S2 is the variance of a sample. X represents each value in the sample (from i=1 to n), while M is the Mean of sample and N is total number of values taken in the sample, for calculating variance. Squaring each value makes the result of subtraction positive, so that the values above Mean do not cancel the values below Mean.

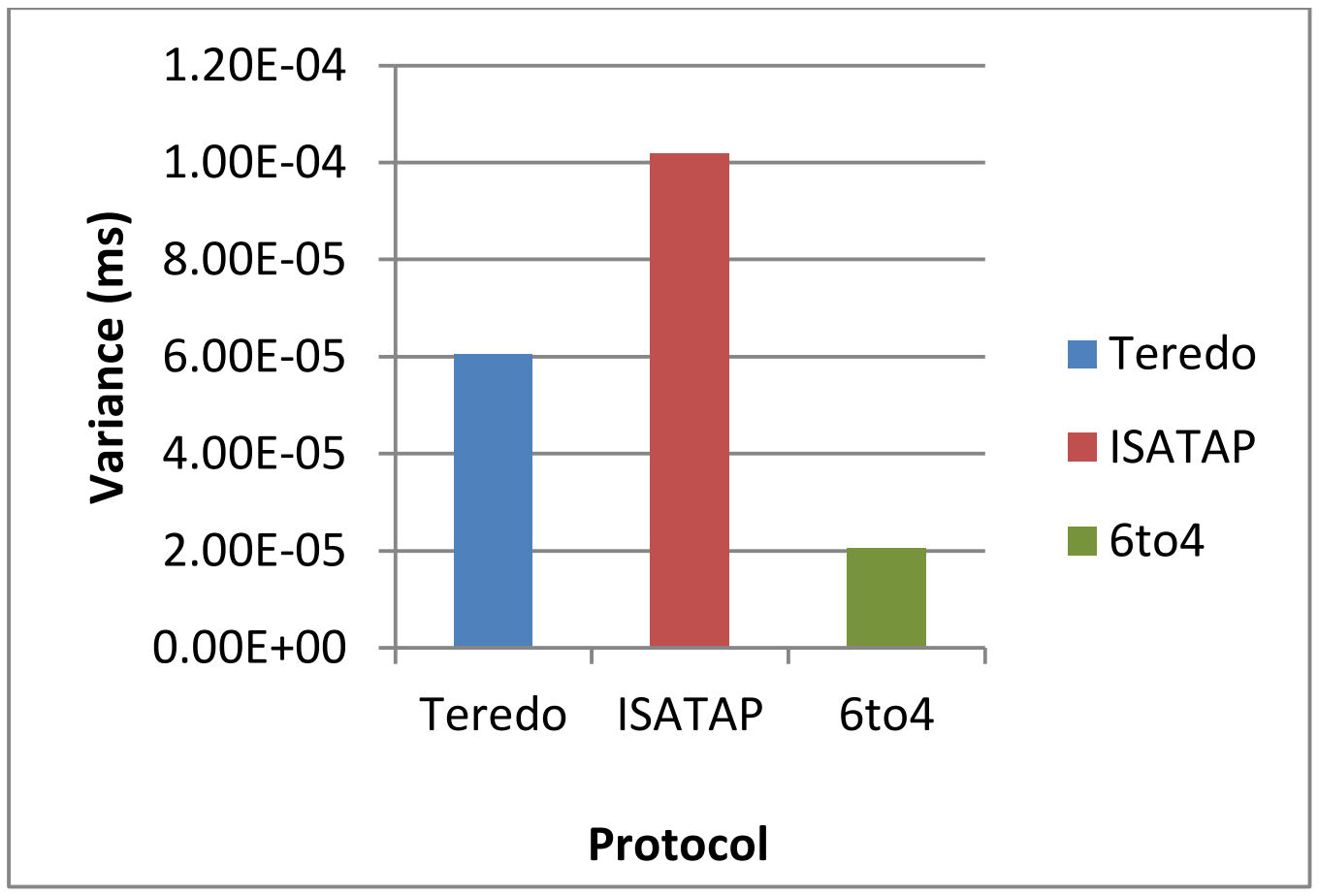

Figure 9 6to4-Teredo-ISATAP Jitter Variance

- Standard Deviation

It is the measure of how much spread out the values are. It is the most commonly used measure of dispersion.

Standard Deviation is simply the square root of variance, shown in equation 6.

$$S = \sqrt{\sum_{i=1}^{N} \frac{(x_i - M)^2}{N}} \qquad (6)$$

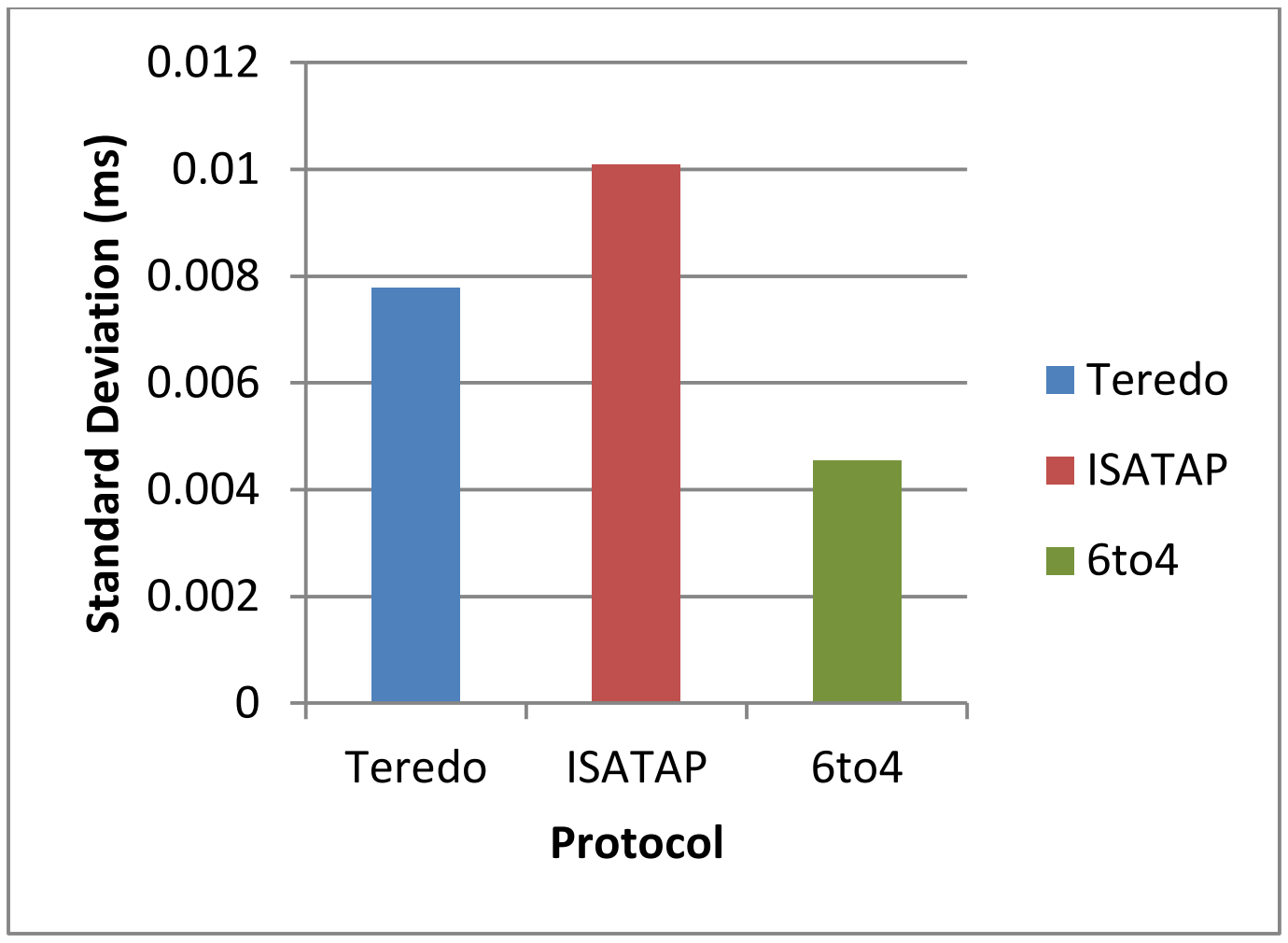

Figure 10 6to4-Teredo-ISATAP Standard Deviation of Jitter

The table IV shows the variance and standard deviation of a 100 packets sample.

TABLE IV. VARIANCE AND STANDARD DEVIATION

| Protocol | Teredo | ISATAP | 6to4 |
|---|---|---|---|
| **Mean Jitter (ms)** | 0.0080 | 0.0290 | 0.0225 |
| **Variance** | 6.0445E-05 | 0.000101775 | 2.05711E-05 |
| **Standard Deviation** | 0.007774638 | 0.010088351 | 0.004535538 |

Variance shows the variability of jitter and hence, provide better picture about jitter here. Since variance is the squared differences between each value and mean value, so it makes every term positive. By this, values above the mean do not cancel out the value below mean. Also, squaring adds more weight to the larger values. Sometimes this can be useful, as values further from the mean may be significant.

*D. Round Trip Time (RTT):* Round trip time is also regarded as one of the key parameters when talking about networks and network layer protocols. RTT is the time taken in total, starting from the moment packet left the sending machine till the reply packet from the receiver is received by the sending machine. To get this, TCP based ICMP-ping traffic was used.
Formula is given in equation 7.

$$RTT = \sum_{i=0}^{n} Tr_i - Ts_i \text{ milliseconds} \quad (7)$$

Where $Tr_i$ represents timestamp of received *i-th* packet. $Ts_i$ represents timestamp of sent *i-th* packet.

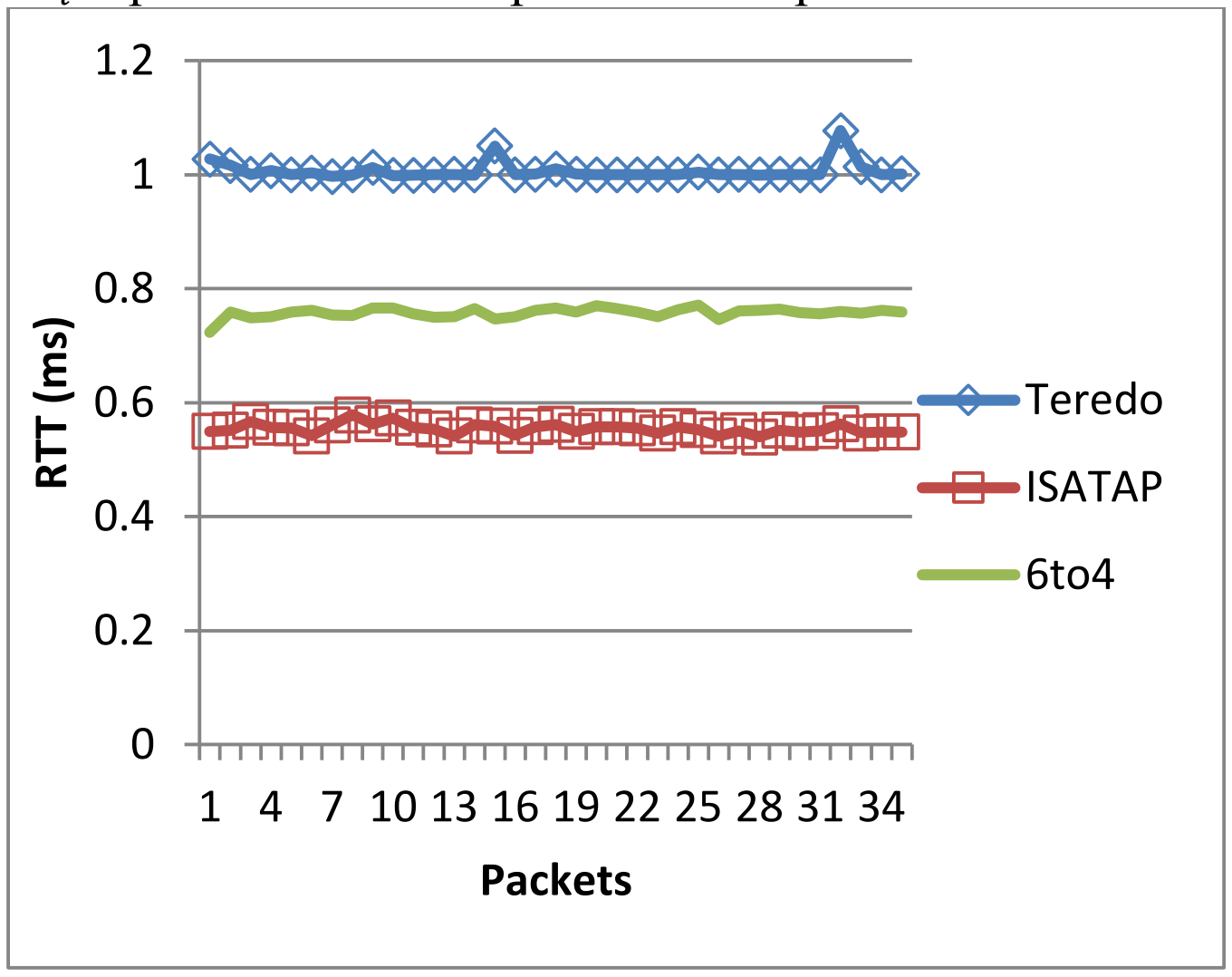

Figure 11 6to4-Teredo-ISATAP round trip time

The difference between E2ED and RTT is that different types of traffic were used to calculate each of them. For E2ED, UDP audio/video streams were used, while for RTT, TCP based ICMP-ping traffic was used.
Average RTT is shown in the table V.

TABLE V. AVERAGE ROUND TRIP TIME

| **Protocol** | **6to4** | **Teredo** | **ISATAP** |
|---|---|---|---|
| **Average RTT (ms)** | 0.7193 | 1.0048 | 0.5516 |

In terms of RTT, ISATAP is again the better choice, because of lesser encapsulation overhead.

*E. Tunneling Overhead:* It refers to the amount of overhead caused by creating tunnels, deleting tunnels, refreshing and maintaining tunnels, encapsulation, and decapsulation.
Tunneling overhead can be calculated by subtracting each protocol's round trip time with the round trip time of native, untunneled/direct traffic. The untunneled or direct traffic was generated by making every node dual stack.. By doing this, exact amount of difference can be calculated between tunneled and untunneled traffic and the overhead caused by sending tunneled traffic can be calculated.

$$Tunneling\ Overhead = RTT_{tt} - RTT_{ut}\ milliseconds \quad (8)$$

Where $RTT_{tt}$ represents round trip time of tunneled traffic and $RTT_{ut}$ represents round trip time of untunneled traffic. For each tunneling protocol, tunneling overhead was calculated separately, using this equation.

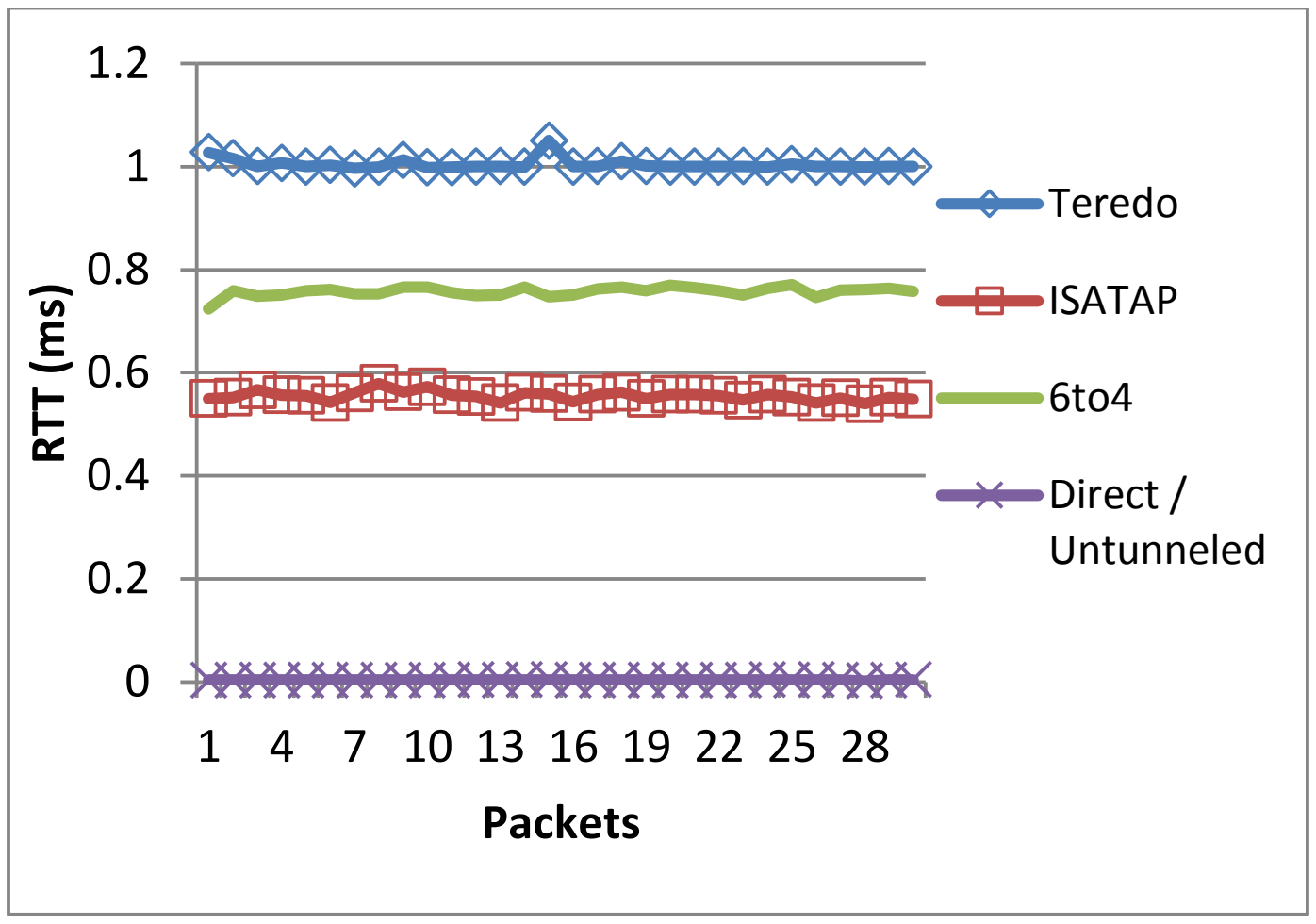

Figure 12 6to4-Teredo-ISATAP Tunneled vs. Untunneled Traffic

Figure 12 shows the difference between tunneled and untunneled traffic. This untunneled traffic is the traffic sent over the same virtual network with all the nodes capable of understanding IPv6. By this, it gives the clear picture of the difference caused by tunneling. Figure 13 gives the exact amount of overall tunneling process, after subtraction.

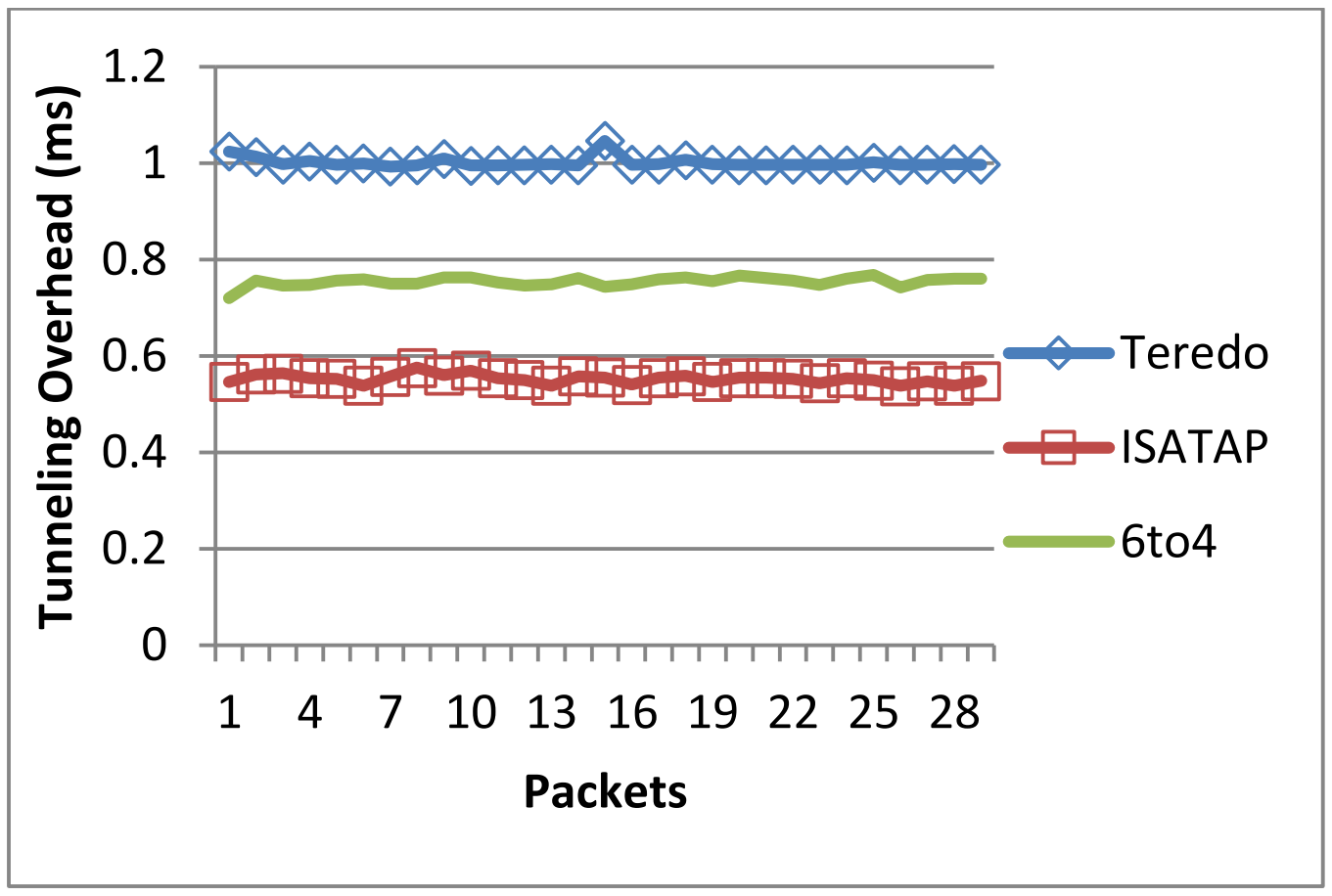

Figure 13 6to4-Teredo-ISATAP Tunneling Overhead

It includes the tunnel setup delay added with processing delay, transmission delay, tunnel refreshing & maintaining, and tunnel teardown. It tells that which tunneling protocol has got how much of tunneling overhead. Average tunnel overhead is shown in table VI.

TABLE VI. AVERAGE TUNNELING OVERHEAD IN MILLISECONDS

| **Traffic** | **RTT (ms)** |
|---|---|
| 6to4 | 0.7124 |
| Teredo | 1.0011 |
| ISATAP | 0.5683 |
| Untunneled | 0.0037 |

As shown in the table V, Teredo's overhead is more than 6to4 and ISATAP, because of its dual encapsulation. As Teredo routers have to encapsulate IPv6 inside UDP-IPv4 headers, so it increases tunneling overhead. Similarly, when the Teredo border router performs decapsulation at the destination end, it has to decapsulate IPv4 and UDP to get the required IPv6 packet that was being tunneled. ISATAP remains the best option in this parameter.

*F. Tunnel Setup Delay:* The total amount of delay incurred in setting up the tunnel to route on the incompatible network. This includes the delay caused by connecting to the tunnel server to get the tunnel prefix and in the end tunnel setup confirmation.

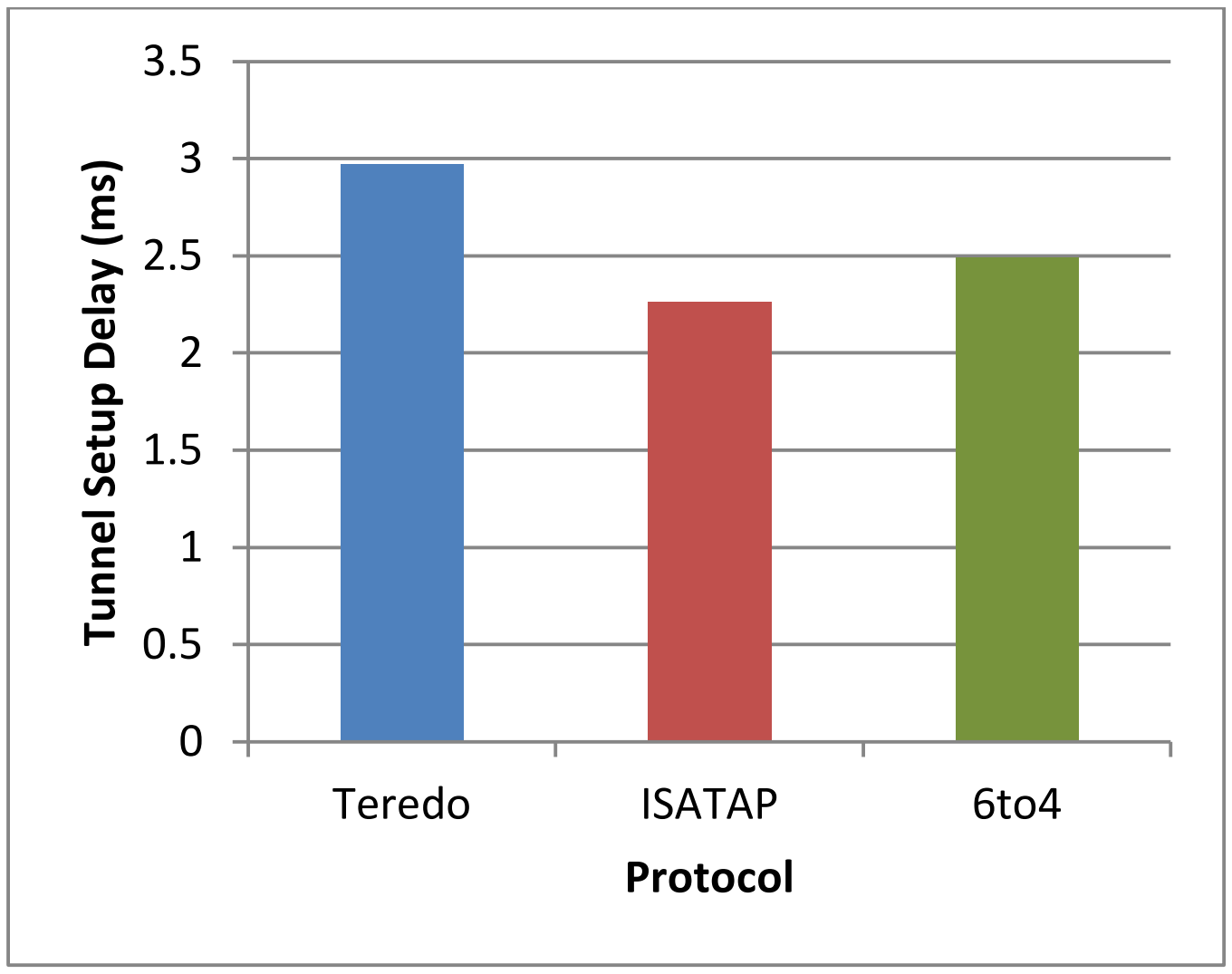

Figure 14 6to4-Teredo-ISATAP Tunnel Setup Delay

Significance of this parameter can be viewed such a way that it tells that how much time it will take to start sending data on an incompatible network. This can be important for addressing quality of service and for critical alerts and signals. Tunnel Setup Delay depends upon the type and size of address prefix to be assigned to the requesting sender node. Other than that, encapsulation overhead also comes into play. It takes 2.49 milliseconds to 6to4 server in setting up the tunnel for its client. After this initial tunnel setup, the data traffic is routed in normal way by the 6to4 routers over the incompatible IPv4 network. Teredo's server takes 2.97 milliseconds for this purpose. The reason for Teredo server being comparatively slow is the same dual encapsulation and different address prefix. Also, the Teredo router does not have the privilege to entertain such queries. It has to pass the requests on to the Teredo server, causing more delay. ISATAP again remains best among the rest, as its server takes 2.26 milliseconds in doing this.  ISATAP setup does not necessarily require a dedicated server. ISATAP router can also perform the same tasks, until for scalability reasons, server is needed.

TABLE VII. AVERAGE TUNNEL SETUP DELAY

| **Protocol** | **6to4** | **Teredo** | **ISATAP** |
|---|---|---|---|
| Tunnel Setup Delay (ms) | 2.49 | 2.97 | 2.26 |

*G. Query Delay:* The delay incurred in querying the DNS server is referred to as query delay. The significance of this parameter is that the tunneling protocols have got different IP prefix, which the DNS has to keep on updating with time. Also, depending upon the scenario, the location of end nodes also matters in this regard. The host may be lying in an IPv6 network. It then has to connect to IPv6-only DNS server to put the query. If the host is lying in an IPv4 network, then the DNS has to be dual stack to maintain the entries of IPv6 hosts lying in IPv6/IPv4 networks, which are being queried. This delay also depends upon the number of hosts querying the DNS server. As the DNS protocol's version, IP prefix, number of querying nodes, and comunicating nodes' location, all can vary, thus, this parameter was considered vital to evaluate which protocol has got more query delay and which has got relatively lesser.

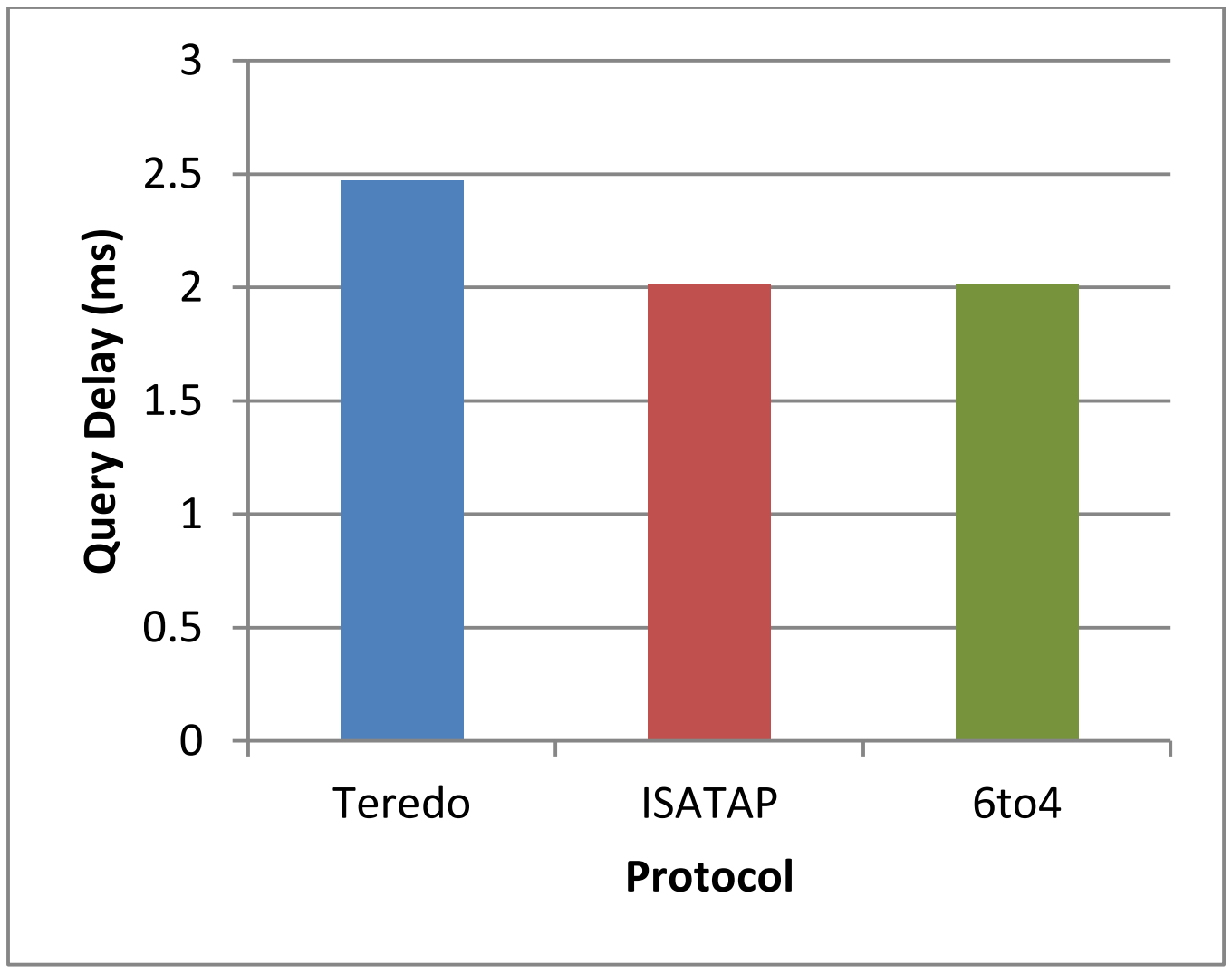

Figure 15 6to4-Teredo-ISATAP Query Delay

In the scenario used for this performance comparison, both, ISATAP and 6to4 have got same average query delay. But as discussed above, this delay can vary, depending upon the locations of end nodes and the DNS server. After multiple independent runs of traffic, the average query delay comes for ISATAP and 6to4 is 2.01 milliseconds, while for Teredo, it is 2.47, relatively higher than the other two. Teredo client is always sitting behind NAT box. So, it also affects the querying process as well.

TABLE VIII. AVERAGE QUERY DELAY

| Protocol | 6to4 | Teredo | ISATAP |
|---|---|---|---|
| Query Delay (ms) | 2.01 | 2.47 | 2.01 |

*H. Auxiliary Devices Required:* The protocols used in this research work involve protocol specific and network specific devices. So, other than the common or by-default devices, the other 'extra' devices are referred to as Auxiliary Devices Required in this study. These devices might not matter in small networks, but they do matter when deployed on a large scale. It all depends upon the type of network (IPv6-only, IPv4-only, dual stack, public network, private network) and the conditions of network. That is why; this parameter has been made part of the evaluation.

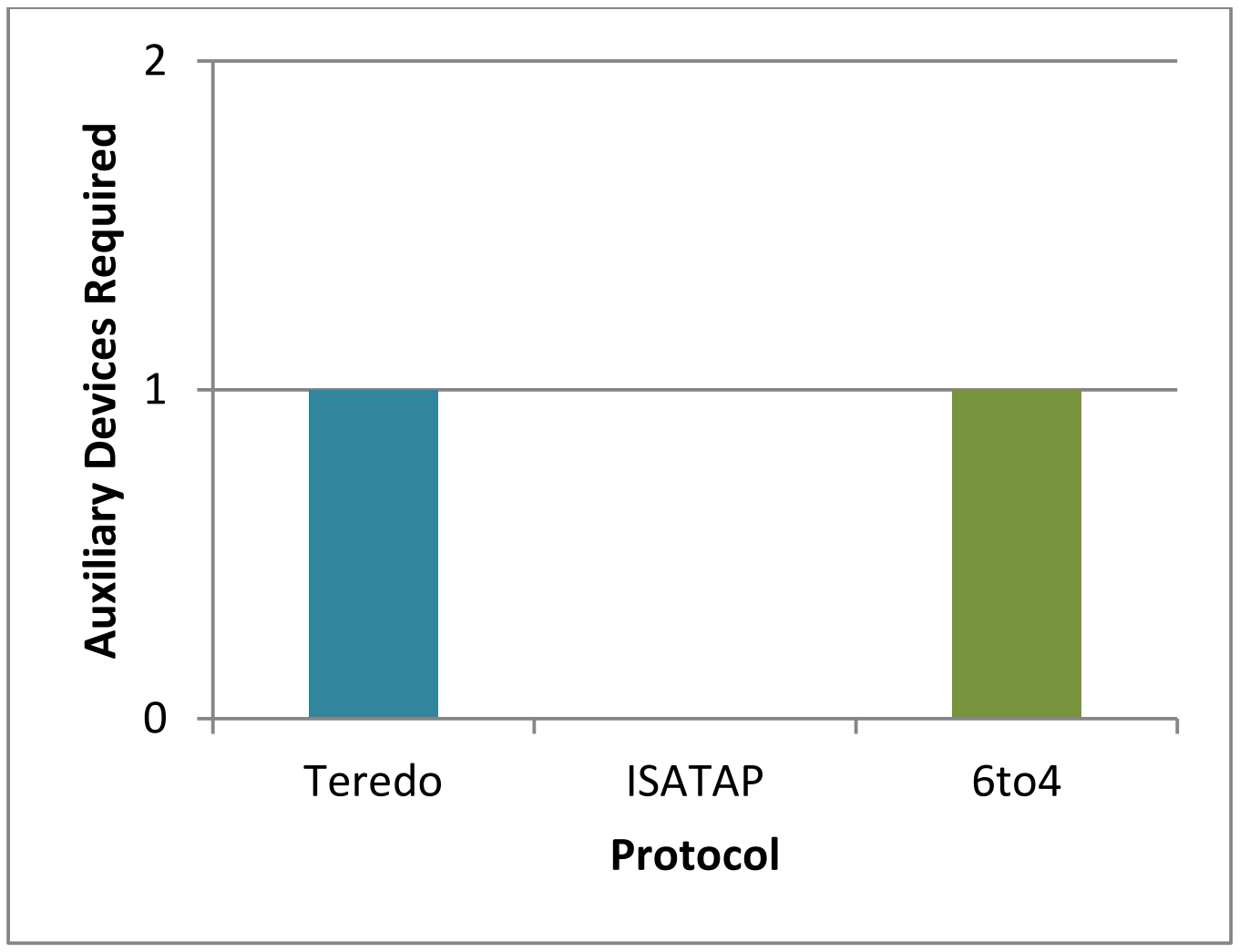

Figure 16 6to4-Teredo-ISATAP Auxiliary Devices Required

6to4 involves an auxiliary device, called 6to4 relay. 6to4's RFC [3] states that the 6to4 relay is basically a 6to4 border router that connects isolated 6to4 routers to the IPv6 Internet. 6to4 tunnels and connections do not require any negotiation with the 6to4 relay from the user or service provider.

In case of Teredo, a standalone Teredo server is required to be deployed to enable Teredo tunneling. Although, a server is required for 6to4 as well, but for smaller networks, that server can also work as a 6to4 router as well. But in Teredo's case, its RFC [17] says that even if the network is small, there should be a separate, standalone server, which would be serving the purpose of initial tunnel setup and later on, tunnel refresh. While the traffic would be routed through Teredo routers, the standalone server allows routers to perform the operation of dual encapsulation and decapsulation and routing, relatively quickly and efficiently.

In case of ISATAP, when the network is small and number of clients who want to send tunneled data through ISATAP are not huge, then there would not be any auxiliary device required. ISATAP's router can do the job of ISATAP server as well, when the functionality is integrated into it. This gives ISATAP an edge over 6to4 and Teredo.

TABLE IX. NO. OF AUXILIARY DEVICES REQUIRED AND DEVICE NAME

| Protocol | 6to4 | Teredo | ISATAP |
|---|---|---|---|
| Auxiliary Devices Required | 1 | 1 | 0 |
| Device Name | 6to4 Relay | Teredo Server | *Not Applicable* |

The table X, given below shows the overall picture of the performance of each protocol, based on all the parameters considered for evaluation in this study.

TABLE X. OVERALL STANDING OF EACH PROTOCOL AGAINST ALL THE PARAMETERS

| **Parameters** | **Rank, in terms of performance** *(1 = comparatively best)* | | |
|---|---|---|---|
| | **1** | **2** | **3** |
| **Throughput** | ISATAP | 6to4 | Teredo |
| **End to End Delay** | ISATAP | 6to4 | Teredo |
| **Jitter** | Teredo | 6to4 | ISATAP |
| **Round Trip Time** | ISATAP | 6to4 | Teredo |
| **Tunneling Overhead** | ISATAP | 6to4 | Teredo |
| **Tunnel Setup Delay** | ISATAP | 6to4 | Teredo |
| **Query Delay** | *Same performance* ISATAP and 6to4 | | Teredo |
| **Auxiliary Devices Required** *+ means none* ** means one* | ISATAP+ | *Same performance** 6to4 and Teredo | |

## V. CONCLUSION

For the tunneling overhead and tunneled traffic comparison of these IPv6 tunneling techniques: ISATAP, Teredo and 6to4; UDP audio streams, video streams, and ICMP-ping traffic were used. In cloud virtualization environment, traffic was run end to end and sniffed at every node for better analysis. Multiple runs of traffic were routed over the network setup for each protocol. At the end, the average of those runs of traffic was calculated to generate final results.

After all this research activity, it is concluded that based on the parameters used: end to end delay, round trip time, throughput, jitter, tunneling overhead, tunnel setup delay, query delay, and auxiliary devices required; overall, ISATAP has got the edge over Teredo and 6to4 in all the parameters, except jitter. ISATAP would be the most efficient and economical solution to tunnel IPv6 packet over IPv4 networks, in most of the situations, when both these versions of IP have to coexist. Although, Teredo has got its own significance, as it would be the best solution for real-time traffic, as it is relatively less jittery. This is because Teredo nodes send tunnel refresh packets less frequently, as compared to ISATAP. But in otherwise cases, ISATAP remains the most suitable choice. 6to4's performance was in between the other two protocols, in every parameter. In cloud computing virtualization, multiple VM's and virtual networks exist in a single server. So the effect of IPv4-IPv6 coexistence does matter a lot. This is what the purpose of this study was, so that, it becomes clear that how much IPv4-IPv6 coexistence impact in cloud virtualization environment.

## VI. FUTURE SCOPE

This work can be extended by doing such activity on WAN environment. The evaluation done in this paper is for virtualization environment in cloud computing. This gives average difference of performances of each protocol and overall evaluation of each protocol. But, if evaluated in Internet-like environment, then it can be shown that how each protocol behaves in unstable conditions and how they will work when actually deployed by service providers in future. Similarly, as discussed in this paper that location of communicating nodes also matter. So, for some particular parameters, like, Query Delay and Tunnel Setup Delay, evaluation in varied scenarios would also be useful. Also, with IPv4-IPv6 coexistence impact, it would be worthy to evaluate that what is the effect of number of virtual machines and resource utilization on the performance of a virtual network.

## ACKNOWLEDGEMENT

This research was supported by the MSIP (Ministry of Science, ICT & Future Planning), Korea, under the ITRC (Information Technology Research Center) support Program (NIPA-2013-H0301-13-1006) supervised by the NIPA (National IT Industry Promotion Agency). Corresponding author: Prof. Eui-Nam Huh.